\documentclass{elsart3p}

 \usepackage{graphics}
 \usepackage{graphicx}

\usepackage{amssymb}

\begin{document}

\begin{frontmatter}



\title{Point-contact spectroscopy of the antiferromagnetic
superconductor HoNi$_{2}$B$_{2}$C}


\author[a]{Yu. G. Naidyuk},
\author[a]{O. E. Kvitnitskaya},
\author[a]{I. K. Yanson},
\author[b]{G.\ Fuchs},
\author[b]{K.\ Nenkov},
\author[b]{A.\ W\"{a}lte},
\author[b]{G.\ Behr},
\author[b]{D.\ Souptel},
\author[b]{S.-L. Drechsler}

\address[a]{B. Verkin Institute for Low Temperature Physics and
Engineering, National Academy  of Sciences of Ukraine,  47 Lenin
Ave., 61103, Kharkiv, Ukraine}
\address[b]{Leibniz-Institut f\"{u}r Festk\"{o}rper- und
Werkstoffforschung Dresden e.V., Postfach 270116, D-01171 Dresden,
Germany}

\begin{abstract}
The point-contact (PC) spectroscopy study of the
electron-phonon(boson) interaction (EP(B)I) spectral function in
HoNi$_2$B$_2$C reveals phonon maxima at 16 and 22\,meV and
34\,meV. For the first time the pronounced high energy maxima at
about 50\,meV and 100\,meV were resolved. Additionally, an
admixture of a crystalline-electric-field (CEF) excitation with a
maximum near 10\,meV and a `magnetic` peak near 3\,meV are
observed. The contribution of the CEF peak in EP(B)I constant
$\lambda_{PC}$ is evaluated as 20-30\%, while contribution of the
high frequency modes at 50 and 100\,meV amounts about 10\% for
each maxima. PC Andreev reflection measurements below the critical
temperature $T_c\simeq$8.5\,K reveals two different
superconducting (SC) states separated by $T_c^{*}\simeq5.6$\,K
which is close to the N$\acute{e}$el temperature
$T_N\simeq$5.3\,K. Below $T_c^*$ the gap $\Delta$ in
HoNi$_2$B$_2$C exhibits a standard single-band BCS-like behavior
with $2\Delta/k_B T_c^{*}\simeq 3.9$.
Above $T_c^*$ the gap features in PC spectra are strongly
suppressed pointing to a peculiar SC state between $T_c^*$ and
$T_c$, where incommensurate magnetic order develops.

\end{abstract}

\end{frontmatter}

By point-contact (PC) researches both the superconducting (SC) gap
and the PC electron-phonon(boson) interaction (EP(B)I) function
$\alpha^2_{\rm PC}F(\omega)$ can be established from the first and
second derivatives of the $I(V)$ characteristic of PC's
\cite{Naid}. Thus the PC spectroscopy is a powerful method to
study both EP(B)I spectra and SC gap behavior.


\begin{figure}
\begin{center}
\includegraphics[width=8cm,angle=0]{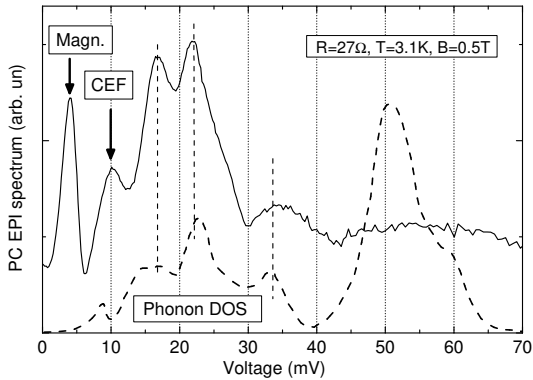}
\end{center}
\caption{PC spectrum of HoNi$_{2}$B$_{2}$C--Cu contact with
subtracted background in comparison  with the neutron phonon DOS
for LuNi$_{2}$B$_{2}$C \cite{Gompf}. Arrows mark position of CEF
peak and `magnetic` peak at about 4\,mV. Vertical dashed lines
mark position of the main phonon maxima in PC spectrum. }
\label{hof1}
\end{figure}

\begin{figure}
\includegraphics[width=8cm,angle=0]{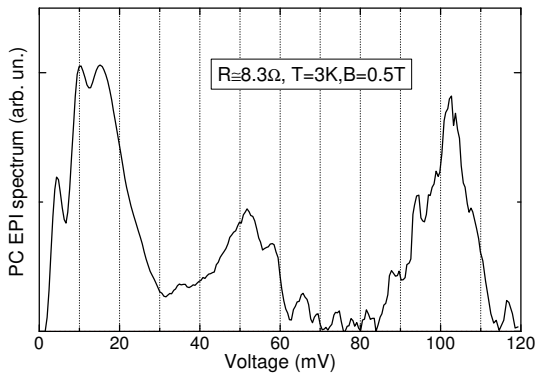}
\caption{PC spectra of HoNi$_{2}$B$_{2}$C--Cu contact with well
resolved high energy maxima around 50 and 100\,mV. } \label{hof2}
\end{figure}

We have measured \cite{NaidSCES} PC spectra of HoNi$_2$B$_2$C with
pronounced phonon maxima at about 16 and 22\,mV, a smeared maximum
near 34\,mV, and shoulder around 50\,mV (Fig.\,\ref{hof1}). All
these features correspond well to the phonon DOS of isostructural
LuNi$_2$B$_2$C \cite{Gompf}, however the high energy part of the
PC spectrum is remarkably smeared. Recently we have also succeed
to measure PC spectra with well resolved high energy maxima around
50 and 100\,mV (see Fig.\,\ref{hof2}). The 50-mV maximum
corresponds to the maximum in the phonon DOS in Fig.\,\ref{hof1},
even shoulder at about 60\,mV is resolved in the PC spectrum. The
phonon maximum at 100-meV was registered for YNi$_{2}$B$_{2}$C in
\cite{Gompf} and it was attributed to the B-C bond stretching
vibrations. The low energy part ($<30$\,mV) of the PC spectra in
Fig.\,\ref{hof2} shows less detailed structure of HoNi$_2$B$_2$C
phonons as compared to the PC spectra from Fig.\,\ref{hof1},
likely due to contribution of the Cu phonons between 15-20\,mV
\cite{Naid}.

The maximum around 10\,mV might be connected with CEF excitations,
observed in this range by neutron scattering \cite{Gasser}, while
the maximum around 4\,mV we attribute to the magnetic order
disturbance (and appearance of spin disorder scattering), since it
disappears above the N$\acute{e}$el ($\sim$6K) temperature. Most
of the PC spectra (not shown) demonstrate expressed 10-mV peak and
completely smeared phonon maxima. This points to the importance of
CEF excitations in the charge transport as well as in the SC
properties of HoNi$_2$B$_2$C. Thus the  contribution of the 10-mV
peak in EP(B)I constant $\lambda_{PC}$ is evaluated as 20-30\%,
while contribution of the high frequency modes at 50 and 100\,meV
amounts about 10\% for each maxima.

\begin{figure}
\begin{center}
\includegraphics[width=8cm,angle=0]{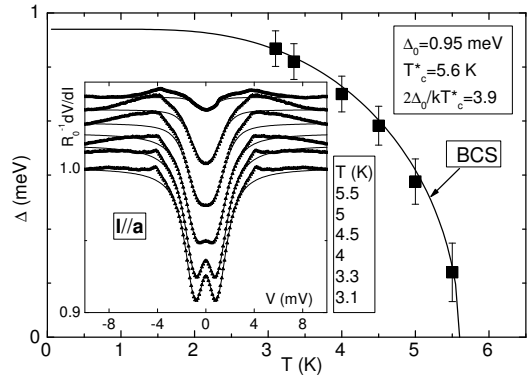}
\end{center}
\caption{$\Delta(T)$ at $B$=0\,T obtained by the fitting of the
curves shown in the inset. Inset: $dV/dI$ curves (symbols) of
HoNi$_{2}$B$_{2}$C--Cu contact ($R=2.7\,\Omega$) established along
a-axis with varying temperature. Solid lines are BTK fitting
curves.}
 \label{hof3}
\end{figure}

The SC gap manifests itself in the $dV/dI$ characteristic of a
N-c-S contact as pronounced minima around $V\simeq\pm\Delta$ at
$T\ll T_{c}$. Such $dV/dI$ are presented in
Fig.\,\ref{hof3}(inset). The SC gap $\Delta$ and its temperature
dependence are obtained (see Fig.\,\ref{hof3}) from the fit of
$dV/dI$ by BTK equations (see \cite{Naid}). It is seen that
$\Delta(T)$ has a BCS-like dependence, however the gap vanishes
already at $T_{c}^{*}\simeq$5.6\,K,  i.\,e. close to the
N$\acute{e}$el temperature $T_{N}\simeq$5.3\,K but well below
$T_{c}\simeq$8.5\,K of the bulk. This can be understood adopting
the ''Fermi surface separation'' scenario for the coexistence of
magnetism and superconductivity in magnetic borocarbides, i.e.\ a
coexistence on different Fermi surface sheets (FSSs), proposed in
Refs.\,\cite{drechsler01}. Thus, we suggest that the
superconductivity in the commensurate antiferromagnetic phase
survives at a special nearly isotropic FSS isolated from the
influence of the rare earth magnetism.

Between $T_{c}^{*}$ and $T_{c}$ the SC signal in $dV/dI$ is
drastically suppressed, giving no possibility to determine a
finite SC gap by fitting the $dV/dI$ data. This shows that the SC
state between $T_{c}^{*}$ and $T_{c}$, in the region of peculiar
magnetic order, is `unconventional`.
We addressed the specific role of selected FSSs and
CEF-excitations in this particular case.

\end{document}